\documentstyle[epsf]{aa}
\newcommand{\Fnu}{\mbox{$F_\nu$}}

\newcommand{\Inu}{\mbox{$I_\nu$}}

\newcommand{\jnu}{\mbox{$j_\nu$}}

\newcommand{\nuz}{\mbox{$\nu_{\rm Z}$}}

\newcommand{\phihat}{\mbox{$\bf\boldmath \hat{\varphi}$}}
\newcommand{\Rein}{\mbox{$R_{\rm{\tiny{E}}}$}}
\newcommand{\rein}{\mbox{$r_{\rm{\tiny{E}}}$}}
\newcommand{\Ds}{\mbox{$D_{\rm{\tiny{S}}}$}}
\newcommand{\Dl}{\mbox{$D_{\rm{\tiny{L}}}$}}
\newcommand{\rsh}{\mbox{$r_{\rm{\tiny{sh}}}$}}
\newcommand{\rhat}{\mbox{$\bf \hat{r}$}}
\newcommand{\rvec}{\mbox{$\bf r$}}

\newcommand{\sini}{\mbox{$\sin i$}}

\newcommand{\vvec}{\mbox{$\bf v$}}

\newcommand{\vz}{\mbox{$v_{\rm Z}$}}
\newcommand{\wobs}{\mbox{$w_{\rm obs}$}}

\newcommand{\zhat}{\mbox{$\bf \hat{Z}$}}
\newcommand{\zstarhat}{\mbox{${\bf \hat{z}}_*$}}

\def \etal{et~al.\/}
\def\lesssim{\mathrel{\hbox{\rlap{\hbox{\lower4pt\hbox{$\sim$}}}\hbox{$<$}}}}
\def\gtrsim{\mathrel{\hbox{\rlap{\hbox{\lower4pt\hbox{$\sim$}}}\hbox{$>$}}}}

\def\figurenum#1{\def\thefigure{#1}\let\@currentlabel\thefigure
\addtocounter{figure}{\m@ne}}
\def\figcaption{\@ifnextchar[{\@xfigcaption}{\@figcaption}}
\def\@figcaption#1{{\def\@captype{figure}\caption{#1}}}
\def\@xfigcaption[#1]#2{{\def\@captype{figure}\caption{#2}}}

\def\fnum@figure{{\rm Figure\space\thefigure:}}
\def\fps@figure{bp}
\def\eps@scaling{.95}
\def\epsscale#1{\gdef\eps@scaling{#1}}
\def\plotone#1{\centering \leavevmode
\epsfxsize=\eps@scaling\columnwidth \epsfbox{#1}}

\begin{document}

\thesaurus{07           % A&A Section 7: Stellar atmospheres
           02.12.3;     % Line: profiles
           08.03.4;     % Stars: circumstellar
           08.09.1;	% Stars: imaging
           08.18.1;     % Stars: rotation
           11.17.2;     % Quasars: emission lines
           12.07.1)}    % Gravitational lensing

\title{Microlensing of Circumstellar Envelopes}

\subtitle{I. Simplified considerations for diagnosing radial and
        azimuthal flow}

\author{R.~Ignace and M.\ A.\ Hendry}

\institute{Kelvin Bldg, Dept of Physics and Astronomy, Univ of Glasgow,
        Glasgow, G12 8QQ, Scotland UK}

\offprints{rico@astro.gla.ac.uk}

\date{Received <date>; Accepted <date>}

\maketitle

\begin{abstract}

This paper presents first results on the line profile shapes from a
circumstellar envelope in bulk motion as modified by a microlensing
event.  Only geometrically and optically thin spherical shells in
uniform expansion or rotation are considered here so as to emphasise
the information content available in the profile shapes.  In particular
it is demonstrated that for the case of expansion, the line emission
can increase by significant factors and the time variation of the
profile shape is {\it symmetric} about line centre.  For uniform
rotation the line emission also increases significantly, but the time
evolution of the profile shape is distinctly {\it asymmetric}.  Thus,
microlensing is seen to yield information about the velocity field in
the extended envelope.  We elaborate on (a) the observational
advantages of tailoring microlensing programs toward detecting extended
circumstellar envelopes, (b) the use of multiline observations to infer
other properties of the envelopes, such as the ionization
stratification, (c) the use of the continuum excess emission at
infrared wavelengths as a means of probing the envelope structure, and
(d) the use of polarisation for constraining the properties of
``clumpy'' winds.


\end{abstract}

\section{Introduction}

In recent years the phenomenon of gravitational microlensing has emerged
as a powerful new tool for analysing the distribution and nature of the
dark matter in the halo of the Milky Way. Since the pioneering suggestion
by Paczy\'{n}ski~(\cite{paczynski1986}) that the systematic monitoring of
several million stars in the LMC and Galactic bulge could be an effective
method for detecting halo dark matter, a number of different monitoring
programs have been established and several hundred candidate microlensing
events detected (Alcock \etal~\cite{alcock+1993},~\cite{alcock+1996};
Aubourg \etal~\cite{aubourg+1993}; Udalski \etal~\cite{udalski+1993},
~\cite{udalski+1994}).

In addition to the detection of dark matter, microlensing studies
have led to significant advances in other areas of astrophysics.
The huge quantity of data from the monitoring programs has
revealed many thousands of new variable stars, prompting important
developments in understanding the physics of Cepheids (Alcock
\etal~\cite{alcock+1995}), RR Lyraes (Alcock \etal~\cite{alcock+1998}),
and eclipsing binaries (Grison \etal~\cite{grison+1995}). In the past
few years several authors have pointed out that the microlensing
events themselves can in principle yield much useful information
about the stars {\it being lensed}, in the case where the stars have
significant angular extent (c.f., Gould~\cite{gould1994}; Simmons \etal\
\cite{simmons+1995b}a; Sasselov~\cite{sasselov1997};
Valls-Gabaud~\cite{valls-gabaud1998}; Hendry \etal~\cite{hendry+1998};
Hendry \& Valls-Gabaud in prep).  Although the majority of candidate
events display light curves which are well-fitted by a simple point
source model, the MACHO collaboration has reported clear evidence of a
microlensing event that exhibits extended source effects (Alcock \etal
~\cite{alcock+1997}). Thus, theoretical consideration of the microlensing
of extended sources has become a timely issue.

The characteristic scale for microlensing is usually taken to be the
Einstein radius of the lens, defined as

\begin{equation}
\Rein = \left [ \frac{4 G M}{c^2} \frac{(\Ds - \Dl) 
        \Dl}{\Ds} \right ] ^{1/2}
\end{equation}

\noindent where $M$ is the lens mass, and $\Ds$ and $\Dl$
are the distance to the source and lens, respectively. In the
microlensing literature it is customary to replace spatial separations
with {\em{angular}\/} separations, conveniently expressed in units of
the angular Einstein radius, $\rein$, namely

\begin{equation}
\rein = \Rein / \Dl = \left [ \frac{4 G M}{c^2} \frac{(\Ds 
	- \Dl)}{\Dl \Ds} \right ] ^{1/2}
\end{equation}

\noindent In the case of a point source lensed by a point mass, the
amplification, $A$, of the source (i.e., the ratio of lensed to unlensed
flux)\footnote{Strictly speaking, since lengths are usually expressed
in terms of angular distance, the expressions derived for the lensed
and unlensed fluxes are often in non-standard units, but 
one is typically concerned with the ratio of lensed to unlensed fluxes,
so that the units cancel for the degree of amplification.} can be shown to
depend only on the {\it projected} angular separation $d$ of the lens
and source, expressed in units of \rein\ of the lens (c.f., Paczy\'{n}ski
~\cite{paczynski1986}; ~\cite{paczynski1996}), namely

\begin{equation}
A(d) = \frac{d^2 + 2}{d\,\sqrt{d^2 + 4}}
	\label{eq:amp}
\end{equation}

\noindent Equation~(\ref{eq:amp}) is independent of the observed
wavelength of the source -- indeed the achromaticity of the light curve
shape is one of the principal defining characteristics of a
microlensing event, usually allowing contamination by variable stars to
be identified in monitoring programs.  However, in the case of an
extended source where the angular stellar radius $R_*$ is
non-negligible as compared to the \rein\ of the lens, the light curve
shape does depend on both the value of $R_*$ and the waveband of
observation (Gould~\cite{gould1994};
Simmons \etal~\cite{simmons+1995b}a; 
Sasselov~\cite{sasselov1997}; Valls-Gabaud~\cite{valls-gabaud1998}). 
This chromatic signature arises principally because
of the wavelength dependence of limb darkening:  the lens ``sees'' a
star with an effective angular radius that changes with wavelength.
Far from being a problem, these extended source effects have been
promoted as a virtue in microlensing studies, since they offer a means
to better constrain the distance, mass and transverse
velocity of the lens (c.f., Simmons \etal~\cite{simmons+1995b}a; 
Alcock \etal~\cite{alcock+1997}).  More
interesting in the present context, however, has been the proposal that
the microlensing of extended sources can be used as a probe of stellar
atmospheres, since the shape of the light curve at different
wavelengths carries information on the radial surface brightness
profile of the source (c.f., Gould~\cite{gould1994};
Simmons \etal~\cite{simmons+1995b}a; Peng~\cite{peng1997};
Valls-Gabaud~\cite{valls-gabaud1998}).

While most authors have considered only the broad-band photometric
light curves produced by extended sources, some recent work has
focussed on the additional information provided by polarimetric and
spectroscopic observations. Simmons \etal~(\cite{simmons+1995b}a) 
and Simmons \etal~(\cite{simmons+1995a}b) 
considered the limb polarisation of
spherically symmetric electron scattering stellar atmospheres as lensed
by a point mass, and showed that maximal polarisations of order one
percent can be achieved during the event -- well within the
capabilities of present-day detection for giant stars at the distance
of the LMC and Galactic Bulge (Newsam \etal\ in prep).
Coleman~(\cite{coleman1998}) improved
this treatment to consider a more general model of a central star
surrounded by an extended spherically symmetric scattering envelope,
with radially varying density.  Newsam \etal~(\cite{newsam+1998}) showed 
that the measurement of polarisation during a microlensing event for such a
model could significantly improve the determination of the stellar
model parameters, compared with model fits based only on broad-band
photometric observations.  Valls-Gabaud~(\cite{valls-gabaud1998}) has
considered the spectroscopic signatures of microlensing from extended
sources, and shows that in addition to providing an unambiguous
signature of microlensing, spectral diagnostics such as the equivalent
width of the Balmer $H\beta$ line can be a powerful means of testing
stellar atmosphere models.

However, a major deficiency in the theoretical treatment of extended
source effects during a microlensing event is the neglect of extended
{\it circumstellar\/} envelopes.  Although the model developed in
Coleman~(\cite{coleman1998}) includes such a scattering envelope, and
Simmons \etal\ (in prep) have further extended this model
to the case where the envelope is aspherical (e.g., a Be~disk), their
computations have been carried out only for the broad-band photometric
and polarimetric response to the event, and the time evolution of line
profile shapes has not been considered.  Although hot stars may be
unlikely sources to be lensed, the line emission from hot star winds
can be quite strong and typically arise from extended regions of
several stellar radii or more, and the amount of emission and the
profile shape depend on the velocity field in the flow.  In this first
paper on the effects of microlensing for circumstellar envelopes, we
employ simplified geometries to investigate the potential of
microlensing for diagnosing the flow structure from the time evolution
of the line profile shapes.

The structure of this paper is as follows. In Sect.~2.1 we derive the
unlensed line profile for a shell that is either expanding or rotating.
Then in Sect.~2.2, we introduce the effects of microlensing,
considering first a direct comparison between the expansion and
rotation cases then investigating the effect of varying the angular
Einstein radius of the lens and the orientation of the lens trajectory.
Finally, in Sect.~3 we discuss implications of our results and future
extensions of our work to utilize multiline observations, continuum
excess emission and polarimetric measurements as further means of probing 
the envelope structure.

\section{Microlensing of Circumstellar Envelopes}

As a first step in describing the diagnostic benefits of
microlensing effects on line profiles formed in circumstellar
envelopes, we consider here only the highly simplified case of a
geometrically and optically thin spherical shell in uniform
expansion or rotation.  The star is also approximated as a point
source of illumination, hence absorption and occultation effects are
ignored.  We first describe the line profiles of these two cases in
the absence of lensing and following derive results that include
microlensing.

\subsection{Unlensed profiles from expanding or rotating shells}

An important consequence of the optically thin assumption is that the
radiative transfer for the line profile reduces to a volume integral
over the envelope, because the assumption implies that every photon
scattered or produced in the envelope escapes.  However, for an
envelope in bulk motion such that the flow speed greatly exceeds the
thermal broadening, the locus of points contributing to the emission at
any particular frequency in the line profile is confined to an
``isovelocity zone'' (c.f., Mihalas~\cite{mihalas1978}).
These zones are determined by the Doppler shift formula, namely

\begin{equation}
\nuz = \nu_0\,\left(1-\frac{\vz}{c}\right),
        \label{eq:doppler}
\end{equation}

\noindent where the observer's coordinates are $(X, Y, Z)$ with the
line-of-sight along $Z$, $\nuz$ is the Doppler shifted frequency, and
$\vz = - \vvec(\rvec)\cdot\zhat$ is the projection of the flow speed
onto the line-of-sight (as indicated in Figs.~1 and 2).  Note that the
stellar coordinates are Cartesian $(x, y, z)$ and spherical $(r,
\vartheta, \varphi)$.  Equation~(\ref{eq:doppler}) will prove crucial for
relating the variable profile shape to the kinematics of the envelope.

For constant expansion with $\vvec = v_0 \rhat$, the velocity shift
along the observer's line-of-sight becomes $\vz = - v_0 \cos \theta$.
For \vz\ constant, the angle $\theta$ is also constant, hence the
isovelocity zone traces a ring on the surface of the shell, as can be
seen in Fig.~1.  For intensity \Inu, impact parameter $p$, and angle
$\alpha$ which is measured from $X$, the observed flux of line emission
from the ring is given by

\begin{equation}
\Fnu = \int\, \Inu(p)\,p\,dp\,d\alpha = 2\pi\,  \Inu(p)\,p\,dp.
	\label{eq:flux}
\end{equation}

\noindent Note again that, in accordance with the standard notation
adopted in the microlensing literature, all distance scales are in fact
taken as angular distances that are normalised to \rein\ of the lens.
This implies that the ``flux'' of Eq.~(\ref{eq:flux}) has rather
unusual units; however, the results of the microlensed line profile
calculations in the following section will be displayed as line ratios
of the lensed to unlensed cases, so that the (non-standard) units of
flux will cancel.

Equation~(\ref{eq:flux}) is not especially instructive as regards the
profile shape.  It is therefore worthwhile to derive an
analytic result for the line profile.  The observed intensity from any
point in an isovelocity ring with impact parameter $p$ is $\Inu(p)
= \jnu(r) dz$, with $\jnu$ the emissivity.  The integral
expression~(\ref{eq:flux}) then becomes

\begin{equation}
\Fnu = \int_{v_{\rm Z}}\,\jnu( r )\, dz\,p\,dp\,d\alpha = \int_{v_{\rm Z}}\,
	\jnu(r)\,dV,  \label{eq:volint}
\end{equation}

\noindent thus a volume integral over the isovelocity zone, as expected
for optically thin line.  In fact using spherical coordinates, the flux
of line emission reduces to

\begin{equation}
\Fnu = \frac{2\pi\,r^2\,\jnu(r)}{v_0}\,dr\,d\vz,
	\label{eq:flattop}
\end{equation}

\noindent where we have substituted $d\mu = -d\vz/v_0$ and integrated
over $d\alpha$ to obtain the factor of $2\pi$.  Since the velocity
shift does not appear in this expression, \Fnu\ is derived to be
constant with frequency, or flat-topped.  The flat-top profile result
in this case is well-known, dating back to Menzel~(\cite{menzel1929}).
Note that the constancy of \Fnu\ in the line implies that $p\,\Inu(p)$
from Eq.~(\ref{eq:flux}) is likewise constant, a result that will be
valuable for numerical calculations of line profiles when microlensing
is included.

Now for the case of uniform rotation, the flow velocity is given by
$\vvec = v_0 \phihat$, for which the velocity shift becomes $\vz = -
\sin \vartheta \cos \varphi \sin i$, where $i$ is the viewing
inclination defined by $\zhat \cdot \zstarhat = \cos i$.  For the
measurement of $\varphi$, we take the observer's $X$--axis to be
coincident with the star's $x$--axis (thus, for a pole-on view, $\alpha
= \varphi$).  Employing spherical trigonometry, it can be shown that
$\cos \beta = \cos \varphi \sin \vartheta$, where $\beta$ is a
spherical polar angle measured from the $X$--axis.  Since the
inclination is fixed for a given shell, the isovelocity zones in the
rotating case once again reduce to circular rings, but now of opening
angle $\beta$ and concentric about the $X$--axis.

Since the isovelocity zone is a ring as in the expanding case, it remains
that the profile shape is flat-topped in the rotating case.
The flux of line emission \Fnu\ for a rotating shell becomes

\begin{equation}
\Fnu = \frac{2\pi\,r^2\,\jnu(r)}{v_0\,\sini}\,dr\,d\vz.
\end{equation}

\noindent The factor \sini\ appearing in the denominator results
from the projection of the shell's rotation axis onto the 
observer's line-of-sight.  Note that for increasingly pole-on
perspectives, the line flux would appear to diverge; however,
an underlying assumption of our derivation is that the thermal
broadening of the line can be ignored.  For nearly pole-on cases,
this assumption is invalid.

It is convenient to define the angular length $q$ as the radius of the
ring, similar to $p$ in the expanding case (see Fig.~2).  Then just as
$p\Inu(p)$ was a constant for an expanding shell, so $q\Inu(q)$ is 
constant for a rotating shell.

We have shown that in the absence of microlensing effects, expanding
and rotating shells both produce flat-top profiles and are therefore
indistinguishable\footnote{Note that for rotation, $v_0$ is limited by
the rotational speed of break-up, whereas expanding winds typically have
terminal speeds significantly in excess of the stellar escape speed.
Hence the line width can provide a distinction between the two cases,
unless the expansion speed happens to be somewhat small.}.  However,
there are two clear distinctions between the expanding and rotating
cases.  (a) In the expanding case, the isovelocity ring is circular in
projection, but for rotation the ring is viewed edge-on and therefore
appears to the observer as a strip of length $2q$.  (b) For expansion
the shell is front-back symmetric, so that for every ring of velocity
shift $\vz$ on the front side of the shell, an exact replica exists on
the back side, only with \vz\ of opposite sign.  However, for rotation
the symmetry is left-right, so that all the points on the left
hemisphere have \vz\ of the same sign, and all the points on the the
right hemisphere also have the same sign but opposite to that on the
left.  These differences in the ring projection and the distribution of
projected Doppler shifts between the two cases will have significant
consequences for the evolution of the line profile shape during a
microlensing event.

\subsection{The effects of microlensing}

\begin{figure}
\plotone{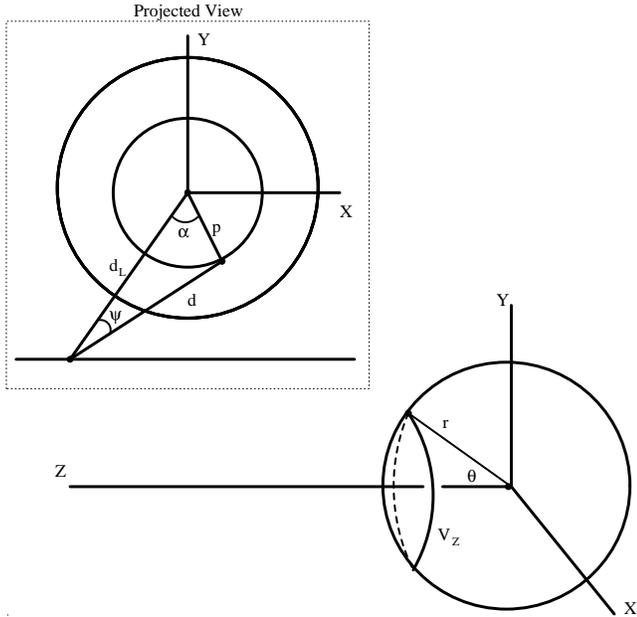}
\caption{Shown is the geometry associated with a uniformly expanding
shell.  Lower left shows the 3-dimensional view with the observer
located on the $Z$--axis.  Circular rings centered on that axis are
distinguished by their Doppler shift $\vz$.  Upper right
(boxed) shows the microlensing event in projection, with the lens a
distance $d_{\rm L}$ from the shell center.   \label{fig:f1}}

\end{figure}

\begin{figure}
\plotone{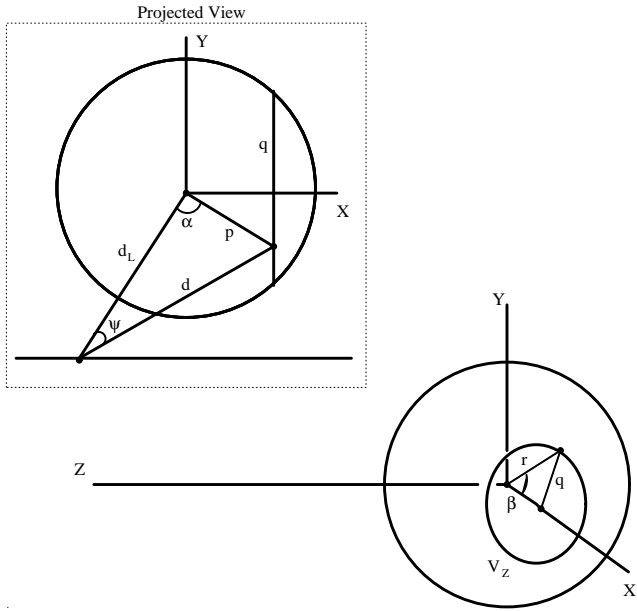}
\caption{Shown is the geometry associated with a uniformly rotating
shell.  The geometry is quite similar to that for an expanding shell,
only now the circular ring is centered on the $X$-axis, with $\beta$
the opening angle of the ring and $q$ the radius of the ring.  In
projection the circular ring is seen edge-on thus appearing as a
vertical strip.  
\label{fig:f2}}

\end{figure}

As discussed in the introduction, the effect of microlensing is to
introduce the amplification factor $A(d)$ from Eq.~(\ref{eq:amp}) into the
flux integral, which now becomes

\begin{equation}
\Fnu = \int\,A(d)\,I(p)\,p\,dp\,d\alpha,        \label{eq:lensed}
\end{equation}

\noindent where $d$ is the projected impact parameter from the lens to
any differential element of material in the envelope.  In
Figs.~\ref{fig:f1} and \ref{fig:f2}, we further define $d_{\rm L}$ as
the impact parameter from the lens to the shell center.  The minimum of
value of $d_{\rm L}$ is denoted by $d_0$ signifying the minimum
impact paramter of the lens trajectory.  Note that Eq.~(\ref{eq:amp})
is the expression derived for the amplification of a point source lens
(c.f., Paczy\'{n}ski~\cite{paczynski1986}), thus
the flux reduces to a convolution of the unlensed intensity and the
amplification by the points lens, integrated across the source plane.
Note that as $d\gg 1$, $A(d)$ tends toward unity, and the unlensed case
is recovered.

Using Eq.~(\ref{eq:lensed}), we have computed emission line profiles
for shells in either uniform expansion or rotation.  Table~\ref{tab:t1}
summarises the simulations that are shown in
Figs.~\ref{fig:f3}--\ref{fig:f7}.  In Tab.~\ref{tab:t1}, the parameter
$\gamma$ defines the orientation of the lens trajectory (c.f.,
Sect.~\ref{subsub:posang}), $A_{\rm max}$ is the maximum amplification
at any frequency in the profile during the microlensing event, and
$f_{\rm line}$ is the maximum enhancement of the {\it total\/} line
emission during the microlensing event relative to that in the absence
of lensing.  The following sections describe the consequence of the
microlensing as it relates to (a) the flow velocity field, (b) the
Einstein radius of the lens, and (c) the orientation of the lens
trajectory with respect to the axis of rotation.

\subsubsection{Expansion versus rotation}

\begin{figure}
\plotone{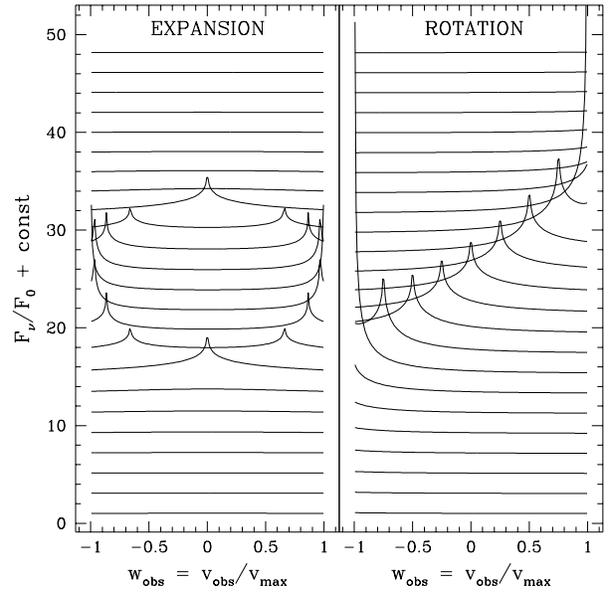}
\caption{Shown are simulations of the microlensing of a uniformly
expanding shell (left) and a uniformly rotating shell (right).  In this
calculation the shell radius is equal to the Einstein radius and the
lens is taken to transit the center of the shell.  Additionally, the
lens trajectory is orthogonal to the projected axis of rotation.  A
constant offset has been applied between the profiles to better show
the time evolution of the line shape (time increasing upwards).  Note
that in the expansion case, the line emission is symmetric about line
center at all times, for rotation the profile is distinctly
asymmetric.  \label{fig:f3}}

\end{figure}

Figure~\ref{fig:f3} contrasts the microlensed line profiles from an
expanding shell (left) to one that is rotating (right), where the
observed velocity shift is $v_{\rm obs}=\vz$ and the maximum velocity
shift $v_{\rm max}$ is equal either to $v_0$ for expansion or
$v_0\,\sin i$ for rotation.  In this simulation we have chosen
$\rsh/\rein=1.0$.  The lens is taken to transit the shell across the
line-of-sight to the shell center with minimum impact parameter
$d_0/\rsh = 0$.  In the case of rotation, we have further assumed that
the lens trajectory is orthogonal to the projection of the rotation
axis in the plane of the sky.  The two panels show a sequence of line
profiles, beginning with the lens at a projected distance of $3 \rein$
from the centre of the shell, which then decreases to zero and
increases back to $3 \rein$ on the opposite side of the shell. Results
are plotted as the ratio of the lensed flux, $\Fnu$, to the unlensed
flux, $F_0$ (the latter being just the flat-top profile), plus a
constant offset introduced between each timestep to better display the
time evolution of the profile shape.

>From Fig.~\ref{fig:f3}, it is evident that the microlensed profiles for
an expanding shell are distinctly different from those of a rotating
shell.  For expansion the profile shape is symmetric both in velocity
about line center and in time with respect to the lens position at
impact parameter $d_0/\rein$.  In contrast, the profile shape in the
rotating case is not symmetric in velocity, and although it is
symmetric in time for the lens trajectory assumed in Fig.~\ref{fig:f3}, the
profile evolution is not time symmetric in general, as will be shown in
Sect.~\ref{subsub:posang}.

The difference between the expansion and rotation cases for the
microlensed line profile evolution can be understood from
considerations of the bulk flow properties of the two cases.  A
sightline from the observer to the lens will intersect the shell at two
points:  one in the front hemisphere and one in the back.  As note
previously, the observed Doppler shift at these two points is equal in
magnitude but opposite in sign for expanding shells, thereby resulting
in an amplification at predominantly two points in the profile, with
the two points being equidistantly located from line center.  However,
for a rotating shell, the Doppler shifts at the two intersection points
are equal both in magnitude and direction, hence the amplification of
line emission occurs predominantly at only one point in the line -- 
resulting in the asymmetric profile shapes.

\subsubsection{Variation of $\rsh/\rein$}

\begin{figure}
\plotone{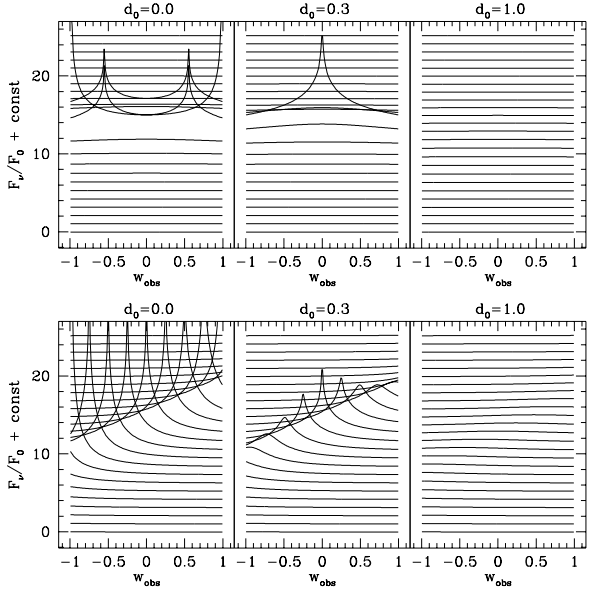}
\caption{With expansion for the upper panels and rotation for the lower
panels, this figure shows the time evolution of the line shape for
$\rsh/\rein = 0.3$.  From left to right, the different cases are for
impact parameters $d_0/\rein$ as indicated.  Note that for this
relatively small value of $\rsh/\rein$ strong microlensing effects
occur, but only for timesteps corresponding to small impact 
parameters.  \label{fig:f4}}

\end{figure}

\begin{figure}
\plotone{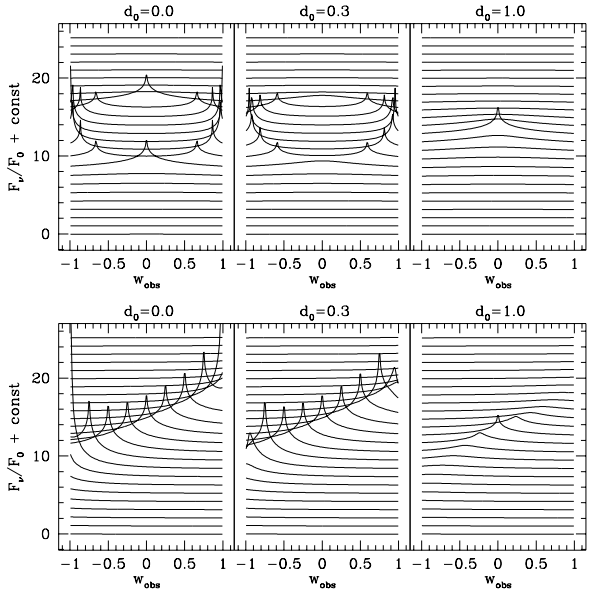}
\caption{As Fig.~\ref{fig:f4} but with $\rsh/\rein=1.0$.  As compared
to the case of $\rsh/\rein=0.3$, interesting microlensing effects now
persist for a broader range of $d_0/\rein$; however, the greatest
amplifications no longer achieve the highest values obtained with the
smaller $\rsh/\rein$ ratio.  \label{fig:f5}}

\end{figure}

\begin{figure}
\plotone{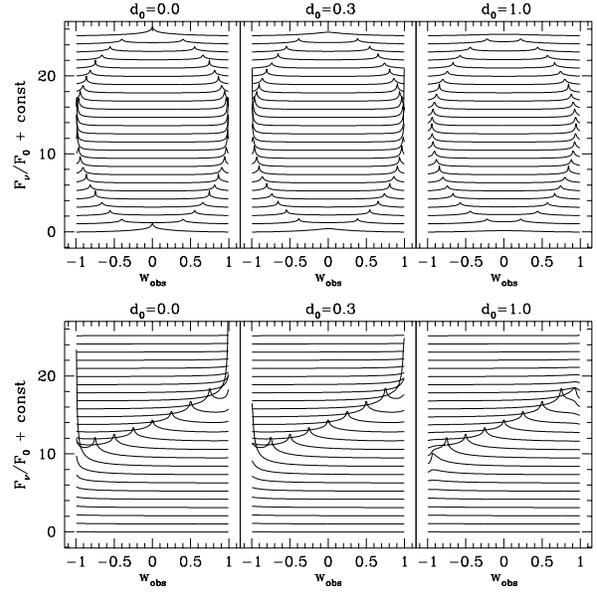}
\caption{As in Figs.~\ref{fig:f4} and \ref{fig:f5}, but for
$\rsh/\rein=3.0$.  Now the effects of microlensing can be seen for quite a
large range of $d_0/\rein$ but the amplifications
are further reduced compared to those for the smaller values $\rsh/\rein$. 
Note also that, since $\rsh$ is large compared with $d_0$, there is little 
to distinguish between results for the three different values of $d_0$. 
\label{fig:f6}}

\end{figure}

Figures~\ref{fig:f4}--\ref{fig:f6} demonstrate how the variation of
the ratio $\rsh/\rein$ affects the response of the line profile
to microlensing.  The three figures are respectively for $\rsh/\rein =$
0.3, 1.0, and 3.0.  Each plot shows six panels, with the upper
set for an expanding shell and the lower one for a rotating
shell.  As labelled, the different panels correspond to different
impact parameters of the lens, with values of $d_0/\rein=$ 0.0,
0.3, and 1.0.  Note that as in the previous section, we assume that 
the lens trajectory intercepts the projected axis of rotation
at a right angle.

Increasing $\rsh/\rein$ leads to three primary effects for the lensed
line profiles:

\begin{itemize}

\item The maximum amplification of the emission line becomes
smaller with increasing values of $\rsh/\rein$.  For more ``compact''
shells, as characterised by smaller $\rsh/\rein$, the amplification is
strongly concentrated to the regions of the shell lying most nearly
behind the lens, which leads to the rather ``spikey'' profiles seen in
Fig.~\ref{fig:f4}.  For sources which are less compact, as in
Figs.~\ref{fig:f5} and~\ref{fig:f6}, the peak amplification is
significantly reduced (refer to $A_{\rm max}$ in Tab.~\ref{tab:t1}).

\item Another interesting effect apparent from Figs.~\ref{fig:f4} --
\ref{fig:f6} is that for small $\rsh/\rein$, significant microlensing
results only for lens trajectories with small $d_0/\rein$, whereas
larger $\rsh/\rein$ show interesting profile effects even for values of
$d_0/\rein \gtrsim 1.0$. This result is more easily understood if we
express $d_0$ in units of $\rsh$.  Consider the rightmost panels of
Fig.~\ref{fig:f4}. Although the minimum impact parameter equals the
Einstein radius, $d_0$ is more than three times greater than $\rsh$,
hence the lens does not actually transit the shell, explaining why an
essentially flat-top profile persists for both expanding and rotating
shells.  Conversely, we see from Fig.~\ref{fig:f6} that significant
lensing is apparent for $d_0 = \rein = \rsh/3$; this is not surprising
since in this case the lens actually transits the shell. The important
conclusion is that significant microlensing is expected in the line
profiles whenever the trajectory of the lens intersects the extended
shell.

\item Fig.~\ref{fig:f6} indicates that for $\rsh/\rein = 3$, there is
little to distinguish between the line profile evolution for the chosen
minimum impact parameters. There are certainly some detailed
differences between the simulations, such as the maximum amplification,
yet the overall time response and average amplification appear quite
similar between the different runs.  This similarity reflects the fact
that $\rsh/\rein$ is relatively large compared to $d_0/\rein$ for these
three cases.  The line profile shape is more sensitive to the value of
$d_0/\rein$ for smaller values of $\rsh/\rein$, as is clear from
Figs.~\ref{fig:f4} and~\ref{fig:f5}.

\end{itemize}

\subsubsection{Variation of the lens position angle}	\label{subsub:posang}

In discussing the position angle, $\gamma$, of the lens trajectory, we
are not concerned with the case of spherically symmetric expanding
shells, for which the microlensing effects are independent of $\gamma$
at fixed $d_0/\rein$.  However, this is not the case for a rotating
shell, because even if the shell density is spherical, the isovelocity
zones in the rotating case are not axially symmetric about the
line-of-sight (except for a pole-on viewing perspective).

\begin{figure}
\plotone{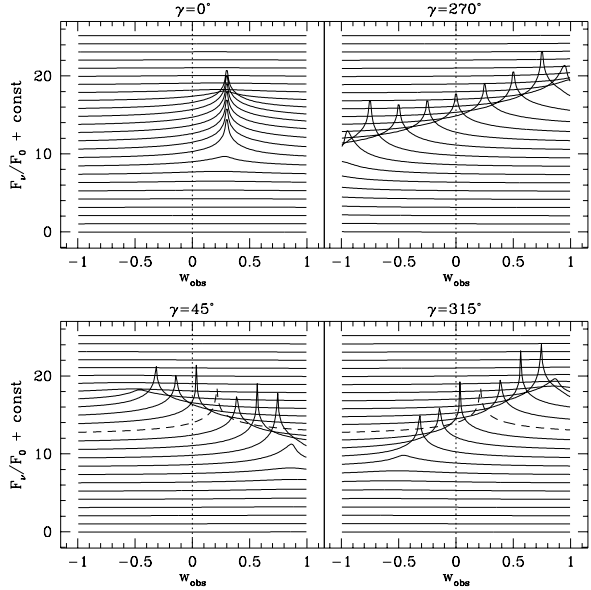}
\caption{Shown are microlensing simulations for a rotating shell with
$\rsh/\rein=1.0$ and $d_0/\rein=0.3$ but allowing the position angle
$\gamma$ of the lens trajectory to vary.  The angle $\gamma$ is
measured counterclockwise from the rotation axis, hence
$\gamma=0^\circ$ is a trajectory parallel to the rotation axis and
$\gamma=270^\circ$ is one that is orthogonal to the rotation axis with
the lens moving left to right.  These simulations indicate that the
relative trajectory of the lens can be recovered from the evolution of
the profile shape.  Note especially that for $\gamma=45^\circ$ and
$315^\circ$, the profile corresponding to the lens' closest approach is
plotted as dashed.  It is evident that evolution of the line profile is
not symmetric in time, a result that is explained more fully in text.
\label{fig:f7}}

\end{figure}

Figure~\ref{fig:f7} shows microlensed line profiles for four values of
$\gamma = 0^\circ$, $45^\circ$, $270^\circ$, and  $315^\circ$.  The
position angle is measured counterclockwise from the $Y$--axis which is
assumed coincident with the projected axis of rotation.  Hence,
$\gamma=0^\circ$ runs parallel to the projected rotation axis in the
direction bottom to top, and $\gamma=270^\circ$ runs orthogonal to that
axis from left to right.  The simulations are for fixed values of
$\rsh/\rein = 1.0$ and $d_0/\rein = 0.3$.

Let us first compare the two cases of $\gamma=0^\circ$ and
$270^\circ$.  Recall that the isovelocity zones are circular rings that
are seen as linear strips in projection, thus for $\gamma=0^\circ$, the
lens trajectory runs along the strip with $d_0/\rsh = 0.3$,
corresponding to $\wobs=0.3$.  Consequently, the peak amplification
always appears at this velocity shift in the profile.  For
$\gamma=270^\circ$, the lens transits each isovelocity strip, and the
result, as in the previous figures, is that the peak amplification
smoothly migrates across the line profile with time.

Now for lens trajectories that are oblique relative to the rotation
axis (as is the case for $\gamma=45^\circ$ and $315^\circ$), the lens
does not transit all of the isovelocity strips.  Neither does the lens
position at minimum impact parameter lie along the isovelocity strip of zero
velocity shift.  The consequence as seen in Fig.~\ref{fig:f7} is that
the peak amplification moves smoothly across the profile but is
significantly greater than unity only for those velocity shifts
where the corresponding isovelocity zones are transitted by the lens.
Further, we have plotted as dashed the line profile corresponding to 
when the location of the lens is $d_0/\rsh$.  The peak amplification at
this time is not at line center.  The evolution of the peak
amplification relative to this time is different for $\gamma=45^\circ$
than for $315^\circ$.  We conclude that the value of $\gamma$ may be
recovered from the profile evolution by virtue of where the peak
amplification occurs in the line profile when the lens is at the 
position of minimum impact
parameter combined with how the peak amplification evolves
across the line relative to that time.

\section{Discussion}

This paper has focussed on the effects of microlensing for the shapes
of emission line profiles from expanding or rotating extended spherical
shells.  From the line profile simulations, it appears that
microlensing, which is normally used as a means of constraining the
properties of the lens, provides a unique and powerful probe of the
stellar source environment.  We find that microlensing can be used
to infer the velocity field of the flow, the ratios $\rsh/\rein$
and $d_0/\rein$, and the orientation of the lens trajectory $\gamma$
in the case of rotation.

Although our treatment has been highly simplified, this work represents
an initial investigation into a broad range of diagnostics for
circumstellar envelope structure from microlensing events.  Here we
summarize avenues of future research on this topic:

\begin{itemize}

\item Previous work on stellar photospheres (c.f., Gould \& Welch
~\cite{gould+welch1996}; Valls-Gabaud~\cite{valls-gabaud1998};
Coleman~\cite{coleman1998}; Hendry \etal~\cite{hendry+1998}) has shown
that microlensing is a unique probe of atmospheric limb darkening,
because the effective stellar radius is a function of wavelength.  In
the same way, the effective radius of circumstellar envelopes is also a
function of wavelength.  The line formation region of some ions is
considerably more extended than for others (see below), and different
continuum emission processes vary in their dependence on parameters
such as density and temperature (also see below). If observed in the
appropriate wavebands, a circumstellar envelope can present a much
larger geometric cross-section than does the stellar photosphere alone.
This is advantageous for the detection of microlensing events, given
that significant amplification will be observed only if the emitting
region lies within \rein\ of the lens.  Consider in particular red
giant stars, which compose  around 20\% of the stars observed by OGLE
in the direction of the Galactic bulge (Loeb \&
Sasselov~\cite{loeb+sasselov1995}), thus comprising a major fraction of
their candidate microlensing events.  Some of these objects have dust
driven outflows, and the dust envelopes produce significant emission at
infrared wavelengths corresponding to geometric sizes of several
stellar radii.  Consequently, microlensing monitoring programs that
target the infrared wavebands should expect a higher incidence of
lensing events as compared to just B~,V or R~bands.

\item We have explored the effects of microlensing only for uniformly
expanding or rotating shells and found that microlensing can provide
valuable information about the velocity field.  For more realistic
extended envelopes (e.g., accelerating winds or Keplerian disks), we
anticipate that microlensing will not only yield the direction of the
flow but also information on the radial derivative.  For example, a
Keplerian disk has slow rotational speeds at large radii and rapid
rotational speeds at small radii.  During a microlensing event, we would
therefore expect the peak amplification to appear first near line
center, migrate toward one extreme wing, then rapidly migrate to the
other wing as the lens moves to the opposite side of the disk, and
finally return to line center as the event concludes.  In contrast, a
very different behavior is anticipated if the rotation were to increase
with radius (e.g., solid body rotation), for which the speed of
rotation is largest at the radius of greatest extent than at the inner
regions.  The line profile would evolve in a manner similar to our
results for uniformly rotating shells, with the peak amplification
appearing first at one extreme wing and migrating across the profile to
the opposite wing.  These two examples assume $\gamma = 270^\circ$, but
similar arguments can be made for other lens trajectories.  The
essential idea is to consider how the lens overlaps different
isovelocity zones as it tracks across the sky.  Microlensing presents a
tremendous opportunity to distinguish between expansion and accretion
and to infer the angular momentum distribution of lensed envelopes.

\item With spectral monitoring during a microlensing event, one could
make relative comparisons of the time evolution in lines of different
ions.  Such measurements allow to determine the ionization distribution
throughout the wind by virtue of the relative time intervals in the
amplifications of different lines.  A natural expectation is that
higher ionization species will exist at smaller radii, closer to the
star, than will lower ionization species.  The lines should have
emission line equivalent widths that are increased by the microlensing
such that the light curves for lines formed at small radii have smaller
FWHM than those formed over a greater radial extent.  A multiline
approach provides a tool of measuring just how stratified the
ionization distribution is.

\item A similar issue is how the continuum emission evolves during a
microlensing event.  There are a variety of continuum emission
processes that may operate in an extended envelope, but at different
spatial locations.  For example, some carbon rich Wolf-Rayet stars show
(a) dust formation that is occuring at large radius (Williams
\etal~\cite{williams+1987}), (b) free-free dominated radio and infrared
excesses over an extended region, and (c) a pseudo-photosphere
dominated by electron scattering at shorter wavelengths.  For such a
star, one expects that during a microlensing event the dust shell is
amplified at far-infrared wavelengths, then amplification of the
continuum emission appears at progressively shorter wavelengths up
until the lens reaches its minimum impact parameter, after which the
amplification subsides in the reverse order.  An especially interesting
aspect of this case is that the amplification at the shortest
wavelengths yields information on the effective stellar radius.  This
radius can in turn be used to measure the relative extent of the
continuum emission throughout the envelope.  The continuum formation
will have some dependence on density (e.g., free-free emission depends
on the square of the density), so that the microlensed data can be used
to infer the radial velocity in the flow.  These kinds of
considerations can also be applied to other types of hot star winds and
to the outflows of evolved late type stars.

\item The previous discussion has implicitly assumed rather smoothly
flowing envelopes, but it is well-known that at least hot star winds
have a considerable level of structure, or ``clumpiness'' 
(Conti~\cite{conti1988}; Henrichs~\cite{henrichs1988}). 
A detailed analysis of the broad band photometric
light curve during a microlensing event may provide information about
the scale and frequency of such clumpy structures, particularly if
$\rein$ is of order the angular scale of the clumps.  Theoretically,
the development of a diagnostic to probe the clumpiness of
circumstellar envelopes bears some similarity to problems involving the
microlensing of photospheres with spots that have been investigated by
Hendry \& Valls-Gabaud (in prep).

\item Lastly, (spectro-) polarimetric observations during a
microlensing event would provide especially detailed information about
the circumstellar envelope structure.  For the purpose of discussion,
consider only polarisation arising from Thomson scattering.  The degree
of polarisation for any single scattering depends on the scattering
angle as $\sin^2 \chi$, hence polarisation is interesting because it
arises primarily from scattering through nearly right angles.  The
convolution of the emergent polarised light with the lens amplification
factor is therefore restricted to spatial regions in the
vicinity of the plane of the sky (i.e., where the scattering angle is
nearly $90^\circ$).  As a result, the polarimetric light curve of a
microlensing event is a sensitive tracer of the radial structure of
the envelope in the plane of the sky.  Such data may provide an even
better diagnostics of clumpiness than the intensity light curve approach
discussed previously.  Moreover, since the number of scatterers depends on
the density (with a linear dependence), the light curve will also provide
information about the radial velocity distribution.

\end{itemize}

There are clearly several tacts for future diagnostic development in
this highly interesting field. Of particular note is the potential
extension of the methods discussed in this paper to the microlensing of
quasars and AGN environments. Observational evidence for quasar
microlensing, based on monitoring the long term variability of broad
band quasar light curves, has been reported by
Hawkins~(\cite{hawkins1993}, \cite{hawkins1996}), although the
microlensing interpretation remains somewhat controversial and several
authors have favoured intrinsic quasar variability as the more likely
paradigm (c.f., Wallinder, Kato \& Abramowicz~\cite{wallinder+1992};
Baganoff \& Malkan~\cite{baganoff+malkan1995}).  Spectroscopic
diagnostics based on modification of both emission and absorption lines
of quasars and AGN during a microlensing event have been investigated
in numerous studies (c.f., Nemiroff~\cite{nemiroff1988}; Schneider \&
Wambsganss~\cite{schneider+wambsganss1990}; Gould \&
Miralda-Escude~\cite{gould+miralda1997}; Lewis \&
Belle~\cite{lewis+belle1998}), which conclude that microlensing can
provide an important probe of density and velocity structure in AGN
environments.  In relation to the results of this paper, Arav
\etal~(\cite{arav+1995}) has applied the line-driven wind theory of hot
stars to interpret the broad absorption lines of quasars, which
indicates that the diagnostic techniques presented here for probing
stellar envelopes may also have relevance to AGN. In particular,
spectral diagnostics may provide an unambiguous signature of quasar
microlensing over a range of lens masses, in which case microlensing
observations may be used to study different components of AGN in
considerable detail.

In conclusion, the techniques, results and discussion presented here clearly 
suggest that microlensing provides a powerful new tool for diagnosing the 
density and velocity structure of extended circumstellar envelopes. We will 
describe the application of these techniques to more realistic stellar models,
and to quasars and AGN, in subsequent papers.

\begin{acknowledgements}

The authors wish to express their appreciation to
Drs.\ J.\ E.\ Bjorkman, N.\ Gray, and D.\ Valls-Gabaud for valuable
discussions on the microlensing of circumstellar envelopes.  This
research was funded by a PPARC rolling grant.

\end{acknowledgements}

\begin{center}
\begin{table}
\caption[]{Lens Parameters for the Model Line Profiles	\label{tab:t1}}
\begin{tabular}{cccccccr}
\hline $\rsh/\rein$ & $d_0/\rein$ & $\gamma$ & \multicolumn{2}{c}{$A_{\rm max}$} & \multicolumn{2}{c}{$f_{\rm line}$} & Figure \\
 & & & Exp & Rot & Exp & Rot & \\	\hline \hline
0.3 & 0.0 & 270$^\circ$ & 33.4 & 83.1 & 5.3 & 3.2 & 4 \\
    & 0.3 & 270$^\circ$ & 13.6 & 9.2  & 5.3 & 2.4 & 4 \\
    & 1.0 & 270$^\circ$ & 1.4  & 1.4  & 1.4 & 1.3 & 4 \\
1.0 & 0.0 & 270$^\circ$ & 10.0 & 36.9 & 1.9 & 2.1 & 3,5 \\
    & 0.3 &   0$^\circ$ & 7.1  & 6.4  & 1.9 & 2.1 & 7 \\
    & 0.3 &  45$^\circ$ & 7.1  & 9.6  & 1.9 & 2.1 & 7 \\
    & 0.3 & 270$^\circ$ & 7.1  & 8.5  & 1.9 & 2.1 & 5 \\
    & 0.3 & 315$^\circ$ & 7.1  & 9.6  & 1.9 & 2.1 & 7 \\
    & 1.0 & 270$^\circ$ & 4.6  & 3.7  & 1.9 & 1.6 & 5,7 \\
3.0 & 0.0 & 270$^\circ$ & 4.5  & 16.0 & 1.2 & 1.5 & 6 \\
    & 0.3 & 270$^\circ$ & 9.4  & 9.1  & 1.2 & 1.5 & 6 \\
    & 1.0 & 270$^\circ$ & 2.3  & 3.7  & 1.2 & 1.4 & 6 \\ \hline
\end{tabular}
\end{table}
\end{center}

\end{document}